\documentclass[english,preprint,showpacs,superscriptaddress,amsmath,amssymb]{revtex4}
\usepackage[T1]{fontenc}
\usepackage[latin9]{inputenc}

\makeatletter

\providecommand{\tabularnewline}{\\}

\@ifundefined{textcolor}{}
{%
 \definecolor{BLACK}{gray}{0}
 \definecolor{WHITE}{gray}{1}
 \definecolor{RED}{rgb}{1,0,0}
 \definecolor{GREEN}{rgb}{0,1,0}
 \definecolor{BLUE}{rgb}{0,0,1}
 \definecolor{CYAN}{cmyk}{1,0,0,0}
 \definecolor{MAGENTA}{cmyk}{0,1,0,0}
 \definecolor{YELLOW}{cmyk}{0,0,1,0}
 }

\usepackage{dcolumn}

\makeatother

\usepackage{babel}

\begin{document}

\preprint{RPC}

\title{Relation Between a Three Parameter Formula for \\
 Istope Shifts and Staggering Parameters}

\author{L. Zamick}

\affiliation{Department of Physics and Astronomy, Rutgers University, Piscataway,
New Jersey 08854, USA }

\date{\today}
\begin{abstract}
It is noted that the staggering parameters used to describe even-odd
effects for isotope shifts can in some cases exhibit very rapidly
varying behavior as a function of neutron number. On the other hand
a three parameter formula (3P) with fixed coefficients can explain
the same behavior. 
\end{abstract}
\maketitle

A three parameter formula was employed by Zamick {[}1{]} and by Talmi
{[}2{]} to describe the isotope shifts in a single j shell $r^{2}(n)=nC+n(n-1)/2A+[n/2]B$
where {[}n/2{]}=n/2 for even n and (n-1)/2 for odd n. Zamick argued
that this should have the same form as a binding energy formula described
in DeShalit and Talmi {[}3{]} because both $r^{2}$ and the two-body
interaction are rotational scalars. We can use this formula to obtain
the staggering parameter which is designed to emphasize the even-odd
behavior of isotope shifts.

The staggering parameter $\gamma$ is defined by

\begin{equation}
\gamma_{n}=2[r^{2}(n+1)-r^{2}(n)]/[r^{2}(n+2)-r^{2}(n)]\end{equation}

One can easily show the relation between $\gamma_{n}$ and the 3P
formula

\begin{equation}
\gamma_{n}=1/(1+M)\ \ \ where\ \ \ M=y/(1+nx)\end{equation}

where y=(A+B)/2C and x=A/C

We thus see that there are only two independant constants for calculating
staggering parameter.That there are only two

is clear from the fact that multiplying A,B and C by the same constant
will not change the staggering parameter although

it will of course change the isotope shift.

%
\begin{table}
\caption{The Staggering Parameters Using the 3P \protect \\
 Formula and A,B,C from Blaum}

\begin{tabular}{c|c|c|c|c}
n  & $\gamma_{n}$  &  &  & \tabularnewline
\hline 
 & Ar  & K  & Ca  & Ti \tabularnewline
0  & 0.388  & 0.660  & 0.052  & -0.387 \tabularnewline
2  & -0.086  & 0.460  & -1.844  & -128.000 \tabularnewline
4  & -3.805  & -4.000  & 3.844  & 2.418 \tabularnewline
6  & 2.908  & 1.853  & 1.948  & 1.705 \tabularnewline
\end{tabular}
\end{table}


By examining Eq. (2) we can quickly get the following results:

a. If A+B=0 then $\gamma_{n}=1$ for all n (no staggering).

b. If A=0 there is no n dependence to the staggering.

c. For very large nA the staggering parameter approaches one. (Since
A is usually very small it is not clear if this limit will be reached
in practice.)

d. Treating n as a continuous variable we can, for small n perform
a Taylor's expansion up to terms linear in n and get

$\gamma_{n}=\gamma_{0}[1+\frac{An}{C}y/(1+y)]$

where $\gamma_{0}=1/(1+y)$.

Since A is usually negative the curve starts on a downslope (e.g.
from Blaum {[}8{]} we have for argon A = -0.018(1), B = 0.119(13),
C = 0.032(5)).

In this work we consider the isotopes of Ar, K, Ca and Ti. Although
the 3P formula was derived for a single j shell, Wolfart et. al. {[}4{]}
showed that it worked if this condition is relaxed. We utilize their
findings plus the fact that the 3P formula has been applied by the
following experimental groups, Andle et al. for Ca{[}5{]}, Martennson-Pendrill
et. al. for K{[}6{]}, Gansky et. al. for Ti{[}7{]} and Blaum et. al.
for Ar{[}8{]}. We follow most closely the work of Blaum et. al. {[}8{]}
where the parameters A, B, C for all these nuclei have been compiled.
From their work we have calculated the staggering parameters shown
in Table 1.

Alternatively, one can get the $\gamma_{n}$'s for Ar isotopes without
the 3P formula. Rather we get the values of $r^{2}(n)$ directly from
Table 3 of Blaum {[}8{]}. However they have no value of $r^{2}$ for
A=45 so we substitute the 3P prediction. For K isotopes we use the
results of A-M Martensson-Pendrill et. al. {[}5{]}.

\begin{table}
\caption{Staggering Parameters for Ar and K Isotopes from Experiment}

\begin{tabular}{c|c|c}
n  & $\gamma_{n}$  & \tabularnewline
\hline 
 & Ar  & K \tabularnewline
0  & 0.527  & 0.274 \tabularnewline
2  & -0.337  & -l.385 \tabularnewline
4  & -3.037  & -0.174 \tabularnewline
6  & 1.461{*}  & 1.278 \tabularnewline
\end{tabular}
\end{table}

{*} Since there is no measurement for 45Ar we use the 3P formula for
this nucleus.

From the definition of the staggering parameters we see that getting
a value close to plus or minus infinity is not so mysterious. One
gets a magnitude of infinity if the A+2 nucleus has the same radius
as the A nucleus. The large negative dips followed by return to positive,
that is displayed in all the above nuclei is easily obtained with
the 3P formula.

Let us discuss the behavior of $\gamma$ vs n for the Argon isotopes.
In a work in which the B(E2)'s and magnetic moments of the even-even
isotopes of these nuclei were calculated by Robinson et. al. {[}9{]}
it was noted that the experimental B(E2)'s rose steadily from A=38
to 42 (midshell for the neutrons) but then steadily decreased for
A=44 and 46. The respective values in units of $e^{2}fm^{4}$ were
130(10), 330(40), 430(100), 345(41) and 196(39). This could explain
why $\gamma$ becomes large and negative for n=4 (which involves 43,44,45
Argon).

The linear term in the 3P formula causes $r^{2}$ to increase with
n. The fact that $^{44}Ar$ has a smaller B(E2) than $^{42}Ar$ would
by itself cause a decrease of $r^{2}$ for $^{44}Ar$ relative to
$^{42}Ar$. If these two effects were to cancel completely,$\gamma$
would have a magnitude of infinity.

In some sense this shows a deficiency in the sheer definition of $\gamma$.
It was designed to emphasize odd-even effects but sometimes the even-even
difference in the denominator can swamp the odd-even difference in
the numerator.

The advantage of the using the staggering parameter is that becuse
it is defined as a ratio some atomic physics parameters that are not
well determined get factored out.In particular there is the F factor
which related the observed frequencies in atomic transitions to isotope
shifts.Thus some experimentalists can measure staggering parameters
with out

being able to measure isotope shifts.Early papers which obtain staggering
parameters include those of Kuhn et. al.{[}10{]}

on tellurium and H.H. Sroke et.al. on mercury isomers{[}11{]}.They
do not discuss the 3P fromula.

The 3P formula has also been applied to heavier nuclei. For example,
Talmi {[}2{]} analyzed the data on lead isotopes by Thompson et. al.
{[}12{]} with this formula. These same isotopes were addressed by
W.H. King et al.{[}13{]} and Anselment et al.

{[}14{]}.These last two authors show that the parmeters used in ref{[}2{]}
show a very flat curve of the staggering parameter

versus neutron number for the lead isotopes in disagreement with experiment.This
is especially shown in a figure

in ref {[}14{]} where it is shown that this parameter decreases rapidly
with decreasing neutron number. One can see,

based on the comments in this work why this is the case. The parameters
used in ref{[}2{]}were

A=\textasciicircum{}-.001, B=0.050, C=0.058. Note that the chosen
A is extremely small.As mentioned before for A=0 the staggering

parameter is independant of n.

The values of Gam found by King {[}13{]} for Isotopes 207 ,205,203,201
and 199 are repectively

0.75, 0.51, 0.45, 0.39, and 0.12 (corresponding to n= 0,2,4,6 and
8). If we fit the first and last of these we obtain

y=.3333 x=A/C=-.1193. In ref {[}2{]} the value of x is -0.0172. With
the new set of parameters the value of GAM

are 0.75, 0.695. 0.61, 0.46, and 0.12. The slope with decreasing n
is much better than in ref {[}2{]} but is far from perfect.

The problem is due to the fact that we are dealing with more than
one shell and more than one value of the ground state spin.

M. R. Pearson et. al. {[}15{]} used the 3P formula to analyze the
bismuth isotopes as well as lead. R. A. Sheline {[}16{]} considered
cesium and barium isotopes. The latter authors discussed anomolous
staggering for which the odd A isotope has an unusually large radius.
They attributed this to octupole deformation which is more prevalent
for odd A as compared to even A. We will not discuss these  nuclei
here but we use them to indicate how widespread the use of the 3P
formula is.

One reason the 3P formula is in fairly wide use by experimentalists
is that more fundamental approaches run into difficulties. Early on
Uhrer and Sorensen {[}17{]} were able, with their pairing plus quadrupole
model, to obtain good results for the even-even to even-even mass
isotope shifts but in their words {}``the odd-even staggering effects
observed in the isotope shifts cannot be obtained''. Sagawa et. al.
{[}18{]} attempted to explain the staggering in lead isotopes by couplings
to giant monopole and quadrupole resonancs. They found that monopole
couplings were much more important but the staggering came mainly
from the quadrupole couplings. They have a nice formula relating the
staggering in isotope shifts to the staggering of B(E2)'s but unfortunately
the staggering is much too small. In the work of Blaum {[}8{]} on
argon isotopes it is noted that spherical Hartree Fock leads to isotope
shifts that are too small. A simulation of deformed H-F increase these
shifts in better agreement with experiment but neither of these approaches
gives odd-even staggering. They imply that an HFB approach might lead
to success {[}8{]}. But their main analysis is with the 3P formula.

We have an added comment on a very recent develpment concerining B(E2)'s
in 46Ar. In ref {[}9{]} we calculate this and compare with values
of Scheit et al{[}19{]}. In units of e$^{2}$fm$^{4}$ Robinson et
al.{[}9{]} obtained 535 as compared with Scheit's value of 196.

A large value was also obtained in the shell model calculations in
the Scheit paper. But a paper just appeared by

Mengoni et al. {[}20{]}where the B(E2) was claimed to be much larger
than the previous measurements. They obtained a value 

of 570 e$^{2}$.fm$^{4.}$This is significantly larger than that for
44Ar 335e$^{2}$fm (for 44Ar good agreement with theory is obtained{[}9{]}

While it is initially gratifying that the new experiment agrees with
the shell model calculations some serious

questions arise when we try to connect with isotope shifts.In ref
{[}8{]} one see that the charge radiuis of 46Ar

is smaller than that of 44Ar. One would expect that if 46Ar has a
bigger deformation than44Ar it should have 

a larger charge radius. It is perhaps premature to drw any definitive
conclusions but it will be interesting to see

how this all plays out.

\section{References$ $}

{[}1{]} L. Zamick, Ann. Phys. 66 (1971) 784.

{[}2{]} I. Talmi, Nucl. Phys. A423 (1984) 189.

{[}3{]} A. de Shalit and I. Talmi, Nuclear Shell Theory, Academic
Press, New York, NY (1963).

{[}4{]} D. H. Wolfart et. al., PRC 23 (1981) 533..

{[}5{]} A. Andle et. al., PRC 26, (1982) 2194.

{[}6{]} A.-M. Martennson-Pendrill et al., J. Phys. B At. Mol. Opt.
Phys. 23 (1990) 1749.

{[}7{]} Yu P. Gangsky et. al., J. Phys. G. Nucl. Part. Phys. 30 (2004)
1089.

{[}8{]} K. Blaum et. al., Nuclear Physics A 799 (2008) 30.

{[}9{]} S. J. Q. Robinson, Y. Y. Sharon and L. Zamick, Phys. Rev C
79 (2009) 04322.

{[}10{]} H.G. Kuhn and R. Turner Proc. R. Soc. Lond. A(1961) 265,39

{[}11{]} H.H. Stroke,D. Proetel and H.J. Kluge, Phys. Lett. 82B (1979)
204

{[}12{]} R. C. Thompson et. al., J. of Phys. G (1983) 443.

13{]} W.H. King and M. Wilson, J. of Physics G: Nucl.Phys. 11(1985)L43

{[}14{]} M.Anselment et. al. NPA 451 (1966) 471

{[}15{]} M. R. Pearson et. al., J. Phys. G Nucl. Part. Phys. (2000)
1829.

{[}16{]} R. A. Sheline et. al., Phys. Rev. C 88 (1988) 2952.

{[}17{]} R. A. Uhrer and R. A. Sorensen, Nucl. Phys. 86 (1966) 1.

{[}18{]} H. Sagawa, A. Arima and O. Scholten, Nucl. Phys. A 474 (1987)
155.

{[}19{]}H.Scheit et al.Phys. Rev. Lett 77,3967 (1996)

{[}20{]} D. Mengoni et. et al. Phys. Rev. C82,04308 2001)
\end{document}